\documentclass[runningheads]{llncs}
\usepackage{graphicx}
\usepackage[T1]{fontenc}
\usepackage{amsmath}

\usepackage{amsthm,amssymb,mathtools}
\usepackage{multicol}
\usepackage{rotating}
\usepackage{soul}
\usepackage{syntax}
\usepackage{graphics}
\usepackage{listings,chngcntr}
\usepackage[dvipsnames]{xcolor}
\usepackage{ifthen}
\usepackage{xspace}
\usepackage{tikz}
\usepackage{stmaryrd}
\usepackage{semantic}
\usepackage{physics}
\usepackage{turnstile}
\usepackage{adjustbox}
\usepackage{ebproof}
\usepackage{rotating}
\usepackage{comment}
\usepackage{utfsym}
\usepackage{xfrac}    
\usepackage{faktor}
\usepackage{sml-highlight}
\usepackage{macros}
\usepackage{semantic}
\usepackage{hyperref}
\usepackage{color}

\usepackage{wrapfig}
\usepackage{svg}
\usepackage{float}
\usepackage{subcaption} 
\usepackage[inline, shortlabels]{enumitem}
\usepackage{ifthen}

\newboolean{extended}
\setboolean{extended}{true} 

\newcounter{listingsCounter}

\begin{document}
\AtBeginEnvironment{lstlisting}{\refstepcounter{listingsCounter}}
\renewcommand{\thelstlisting}{\arabic{listingsCounter}}

\title{A Formal Modeling Language for\\ Smart Contracts}

\author{Adele Veschetti\inst{1}\orcidID{0000-0002-0403-1889} \and
Richard Bubel\inst{1} \and\\
Reiner H\"ahnle \inst{1}\orcidID{0000-0001-8000-7613}}
\authorrunning{A. Veschetti et al.}
%
\institute{Department of Computer Science, TU Darmstadt, Germany\\
\email{\{adele.veschetti,richard.bubel,reiner.haehnle\}@tu-darmstadt.de}}

\maketitle

\begin{abstract}
  Smart contracts codify real-world transactions and automatically
  execute the terms of the contract when predefined conditions are
  met.  This paper proposes \smartML, a modeling language for smart
  contracts that is platform independent and easy to comprehend.  We
  detail its formal semantics and type system with a focus on its
  role in addressing security vulnerabilities. We show along a case study,
  how \smartML contributes to the prevention of reentrancy attacks, 
  illustrating its efficacy in reinforcing the reliability and security 
  of smart contracts within decentralized systems.
\end{abstract}

\section{Introduction}
\label{sec:introduction}

Distributed ledger technologies are realized as a peer-to-peer
network, where each node independently maintains and updates an
identical record of all transactions, known as a ledger. To establish
consensus on the accuracy of a single ledger copy, a consensus
algorithm is employed.  The most popular design for distributed
ledgers employs blockchains, which are immutable lists with built-in
integrity and security guarantees. These assurances, coupled with a
consensus algorithm, establish the trustworthiness of distributed
ledgers.

A central aspect for the usefulness of blockchains is their capability
to store programs, so-called smart contracts, and their (dormant)
runtime state in between transactions. Smart contracts formalize
agreements between parties, such as resource exchange protocols, with
the expectation of providing tamper-proof storage for
security-critical assets. Despite this promise, widely-used smart
contract languages are often complex, rendering them susceptible to
unforeseen attacks. At the same time, in blockchain systems,
rectifying errors post-transaction is nearly impossible. Both aspects
together make it of paramount importance to ensure the correct
functionality of smart contracts \emph{before} deployment.  The fact
that correctness of smart contracts is relevant in practice
is substantiated by the vast number of security vulnerabilities
\cite{AbdellatifB18} and (partially successful) attacks such as DAO
\cite{DAO}, the latter causing damage of 50 million USD worth of
Ether.

In consequence, there is a strong motivation for formal specification
and verification of smart contracts, despite the effort involved, with
a variety of approaches
\cite{AbdellatifB18,verisol,flint,securify}. However, to be feasible
and economical in practice, the effort for specification and
verification of smart contracts must be as small as possible. This is
the focus of our work.

We propose \smartML, a language-independent modeling framework for
smart contracts, that permits to formally prove and certify the
absence of important classes of attacks with a high degree of
automation.
Operating at a high abstraction level, this modeling language is
designed to be easily comprehensible, facilitating the validation of
smart contract behavior. \smartML incorporates abstractions of key
concepts underlying various distributed ledger technologies.

We equip \smartML with a formal semantics to provide a precise and
unambiguous definition of the its meaning and behavior.  The formal
\smartML language is a basis for various kinds of sound analyses,
including type checking, static analysis, and deductive verification.
In the present paper, we showcase a sound type system to prevent
reentrancy in smart contracts by regulating the flow of interactions
between functions. In consequence, by type checking we are able to
avoid unintended recursive calls, reducing the risk of reentrancy
vulnerabilities, but still permitting safe reentrant calls. This
achieves increased overall security of smart contracts by maintaining
strict control over their execution flow without being overly
restrictive and limiting functionality.

The paper is organized as follows. Section \ref{background} contains
an overview of smart contracts and reentrancy attacks to make the
paper self-contained.  The {\smartML} modeling language, along with
its semantics, is detailed in Section
\ref{smartml-lan}. The formal definition of reentrancy security is given in Section~\ref{sec:formalisation-reentrance-safety}, while Section~\ref{sec:assert-safe-reentr} presents the
type system for safe reentrancy. Section \ref{evaluation} provides
examples to illustrate the discussed concepts.  A comparative analysis
of our proposal with existing literature is in Section
\ref{related-works}, while Section \ref{conclusions} summarizes key
findings and explores potential avenues for future research.

\section{Background}
\label{background}


\subsection{Smart Contracts}
\label{sec:smart-contracts-back}

Blockchain technology facilitates a distributed computing
architecture, wherein transactions are publicly disclosed and
participants reach consensus on a singular transaction history,
commonly referred to as a ledger~\cite{bashir2017mastering}. 
Transactions are organized into blocks,
timestamped, and made public. The cryptographic hash of
each block includes the hash of the preceding block, creating an
immutable chain that makes altering published blocks highly
challenging.

Among the applications of blockchains, smart contracts stand
out. These automated, self-executing contracts redefine traditional
agreements, offering efficiency and transparency in various
industries.  A smart contract is essentially a computer program
delineated by its source code. It has the capability to automatically
execute the terms of a distinct agreement expressed in natural
language if certain conditions are met. Typically crafted using
high-level languages, smart contracts are then compiled to bytecode
and encapsulated in self-contained entities deployable on any node
within a blockchain.

Smart contracts can be developed and deployed on various blockchain
platforms, such as NXT~\cite{NXT}, Ethereum~\cite{ethereum}, and
Hyperledger Fabric~\cite{hyperledger}. Each platform offers distinct
features, including specialized programming languages, contract code
execution, and varying security measures.

\subsection{Reentrancy}
\label{sec:reentrancy}

One of the most common vulnerabilities of smart contracts is the
reentrancy attack: it initiates a recursively called procedure
facilitating the transfer of funds between two smart contracts, a
vulnerable contract $\mathcal{C}$ and a malicious contract
$\mathcal{A}$.  The attacker places a call to the vulnerable contract
with the aim of transferring funds to $\mathcal{A}$.  Contract
$\mathcal{C}$ verifies whether the attacker possesses the requisite
funds and, upon confirmation, proceeds to transfer the funds to
contract $\mathcal{A}$.  Upon receipt of the funds, contract
$\mathcal{A}$ activates a callback function, which subsequently
invokes contract $\mathcal{C}$ again \emph{before the balance update
  occurs}.
Different types of reentrancy attacks can be categorized into three
forms, each with distinct characteristic:
\begin{enumerate}
\item \textit{Single Reentrancy Attack}: Here the vulnerable function
  that the attacker recursively calls is the same as the one being
  exploited.
\item \textit{Cross-function Attack}: These occur when a vulnerable
  function shares state with another function that yields a desirable
  outcome for the attacker.
\item \textit{Cross-contract Attack}: It takes place when the state
  from one contract is invoked in another contract before it is fully
  updated.
\end{enumerate}

Specifically, cross-function reentrancy refers to a vulnerability in
smart contracts where an external call is made to another function
within the same contract before the completion of the first function's
state changes.  In other words, during the execution of one function,
an external call is initiated to a \emph{different} function within
the \emph{same} contract, potentially leading to unexpected or
malicious behavior. Identifying this form of reentrancy attack is
challenging: In complex protocols many combinations occur, making it
practically impossible to manually test every possible outcome.
Spotting cross-contract vulnerability is also challenging, because it
involves interactions between multiple smart contracts, making it hard
to foresee the execution sequence and potential vulnerabilities.

One way to avoid certain reentrancy vulnerabilities is to adhere to
the Checks-Effects-Interactions Pattern \cite{WohrerZ18}. This
approach suggests that a smart contract should initially perform
necessary checks (Checks), subsequently modify its internal state
(Effects), and \emph{only then} interact with other smart contracts,
some of which could be potentially malicious. By following this
pattern, a reentrant call becomes indistinguishable from a call
initiated after the completion of the initial call. However, while the
Checks-Effects-Interactions pattern is a crucial guideline for
avoiding reentrancy vulnerabilities within a \emph{single} contract,
it may not provide sufficient protection against cross-contract or
cross-function reentrancy attacks. The inherent complexity of these
attacks, combined with the nature of smart contract interactions,
requires a more comprehensive approach.

\section{\smartML}
\label{smartml-lan}

In \smartML, a program is a sequence of contract declarations and
algebraic data type (ADT) definitions. The combination of contract and
immutable ADT declarations provides a flexible and suitably abstract
approach to smart contract modeling.

\begin{wrapfigure}[10]{r}{.34\textwidth}
\centering\vspace*{-2.9em}
\begin{lstlisting}[style=sml-color, columns=flexible, tabsize=2,basicstyle=\sffamily\small,label={example-code},caption={\smartML Code},captionpos=b]
contract C {
 int n ;
 constructor(int val) { 
  this.n = val; 
 }
 int m(int x) { 
  return n+x; 
 }
}
\end{lstlisting}
\end{wrapfigure}

The selection of language features of \smartML resulted from a
comprehensive, detailed, and systematic analysis of the commonalities
and differences among existing smart contract languages.
Putting a strong focus on security through static analysis and
deductive verification, \smartML employs a formal semantics to ensure
the unambiguous meaning of its operations. The modeling language is
also equipped with a type system designed to address crucial security
aspects like safe reentrancy. Listing~\ref{example-code} shows a
simple smart contract written in \smartML.
    
\subsection{Syntax}
\label{sec:syntax}

The grammar of \smartML is shown in Table~\ref{syntax}. We use the
following syntactic conventions: $C$, $D$ refer to contract names;
$A$, $L$, $K$, $M$ represent ADT declarations, contract declarations,
constructor and method declarations, respectively; $n$ stands for ADT
function declarations; $d$ indicates ADT expressions; $f$, $g$ denote
fields; $m$ stands for method declarations; $s$, $e$, $v$, and $\tau$
cover statements, expressions, values, and types, respectively, while
local variables are denoted by~$x$. We use the overline symbol
($\overline{f}$) to represent a (possible empty) sequence of elements
$f_1,\ldots,f_n$ and square brackets $[~]$ indicate optional elements.

The set of all local program variables is called
$\mathsf{ProgVars}$. For ease of presentation we assume that each
local program has a unique name.  The set of program variables
$\ProgVars$ includes the special variable~\codeinline{this}$_{C}$ for each
contract type $C$. If the context is clear the subscript $C$ is
omitted.

\smartML programs are a series of ADT and contract declarations.
An ADT definition consists of a sequence of function
definitions. Permitted ADT expressions include
\codeinline{if-then-else}, \codeinline{return},
\codeinline{switch}-constructs, and function calls.

\begin{table}[h!]
  \footnotesize
  \[
    \begin{array}{r@{\;}ll}
      P \Coloneqq & \overline{A} \; \overline{L} & \text{(program)}\\
      A \Coloneqq & \mbox{\codeinline{datatype}}\; adt\; \{\mbox{\codeinline{constructor}} \{\overline{fn} :: \overline{\adt}\} \;\overline{F}\} & \text{(datatype)} \\
      F \Coloneqq & \tau_n \; n(\overline{\tau}\;\overline{w})\ \{ d \} & \text{(ADT function)}\\
      d \Coloneqq & \mbox{\codeinline{if}}\; (c)\; \{\; e \;\}\; \mbox{\codeinline{else}}\; \{\;e\;\} \mid \mbox{\codeinline{return}}\;e \mid  n(\overline{w}) & \text{(ADT expression)}\\
                  & \mid \mbox{\codeinline{switch}}\; e \;\{\mbox{\codeinline{case}}\; e : \; d;\; \mbox{\codeinline{default}}: \;d\} & \\
      L \Coloneqq & \mbox{\codeinline{contract}}\ C \ [\mbox{\codeinline{extends}}\ D]\ \{\overline{f} : \overline{\tau}_f; K; \overline{M}\} & \text{(contract)}\\
      K  \Coloneqq & \mbox{\codeinline{constructor}}([\overline{g} : \overline{\tau}_g,]\ \overline{f} : \overline{\tau}_f)\ \{[\mbox{\codeinline{super}}(\overline{g});]\ \mbox{\codeinline{this}}.\overline{f} = \overline{f} \} & \text{(constructor)}\\
      M  \Coloneqq & \tau_m \; m(\overline{\tau}\;\overline{v})\ \{ s \} & \text{(method)} \\
      s  \Coloneqq & \mbox{\codeinline{if}}\; (e)\; \{s\}\; [\mbox{\codeinline{else}}\; \{s\}] \mid \mbox{\codeinline{while}} (e)\ \{s\} \mid \mbox{\codeinline{let}}\; v \coloneq \rhs\; \mbox{\codeinline{in}}\; s  & \text{(statement)}\\
                  & \mid \mbox{\codeinline{assert}}(e) \mid v \coloneqq \rhs \mid v \coloneqq v.m(\overline{v}) \mid v.m(\overline{v}) \mid \mbox{\codeinline{return}}\;e \\
                  & \mid \mbox{\codeinline{throw}}\;e \mid \mbox{\codeinline{try}}\; s_0 \; \mbox{\codeinline{abort}} \; \{s_1\}\; \mbox{\codeinline{success}}\;\{s_2\} \mid s_1 ; s_2\\
      e \Coloneqq & v  \mid e_1 \;\mathtt{op}\; e_2  \mid e_1 \;\mathtt{bop}\; e_2  & \text{(expression)}\\
      v \Coloneqq & x \mid\; !v \mid v.f \mid \mbox{\codeinline{true}} \mid \mbox{\codeinline{false}} \mid d \mid \mbox{\codeinline{this}}_C & \text{(value)}\\
      \rhs \Coloneqq & e \mid \mbox{\codeinline{new}}\;C(\overline{v}) & \text{(right-hand side)}\\
      \mathtt{op} \Coloneqq & + \;\mid\; - \;\mid\; \times \;\mid\; \divisionsymbol  & \text{(arithmetic operator)} \\
      \mathtt{bop} \Coloneqq  & \leq \;\mid\; \geq \;\mid\; \&\& \;\mid\; \| \; \mid \; = \; \mid \; \neq & \text{(boolean operator)} \\
    \end{array}
  \] 
  \caption{The syntax of \smartML.}
  \label{syntax}
\end{table}

A contract declaration
introduces a contract $C$, which may extend a
contract~$D$. Contract~$C$ has fields~$\overline{f}$ with
types~$\overline{\tau_f}$, a constructor~$K$ and
methods~$\overline{m}$.  The set of all contract types (names) is
called $\Contract$, the set of all fields is called $\Field$.
The constructor
initializes the fields of a contract~$C$.  Its structure is determined
by the instance variable declarations of~$C$ and the contract it
extends: the parameters must match the declared instance variables,
and its body must include a call to the super class constructor for
initializing its fields with parameters~$\overline{g}$. Subsequently,
an assignment of the parameters~$\overline{f}$ to the new fields with
the same names as declared in~$C$ is performed.
A method declaration
introduces a method named $m$ with return type $\tau_m$ and parameters
$\overline{v}$ having types $\overline{\tau}$. Most statements are
standard; for instance, $v \coloneqq \rhs$ and
$v \coloneqq v.m(\overline{v})$ denote assignments, and
\codeinline{assert(e)} ensures that a condition holds: If the expression
$e$ evaluates to \codeinline{true}, executing an assertion is like
executing a skip. On the other hand, if the expression evaluates to
\codeinline{false}, it is equivalent to throwing an error.
Expression syntax is standard, however, for field access $v.f$, we
restrict $v$ to \codeinline{this}$_C$. For ease of presentation, we assume that
method invocations pass only local variables as arguments.

This article focuses on contract and statement
definitions. Consequently, we do not detail the semantics and type
system related to ADT definitions.

\subsection{Semantics}
\label{sec:semantics}

We describe the semantics of \smartML in the style of structural
operational semantics (SOS)~\cite{Plotkin04a}. SOS defines a
transition system whose nodes are configurations that represent the
current computation context, including call stack, memory, and program
counter. The SOS rules define for each syntax element of a programming
language its effect on the current configuration.  The general schema
of an SOS rule is
{\small\[
\mathruleW{rule name}{\text{conditions}}
    {\mathit{cfg}[\mathit{stmnt}] \leadsto \mathit{cfg}'[\mathit{stmnt}']}
\]}

It relates a start configuration $\mathit{cfg}$ with the configuration
$\mathit{cfg}'$ reached when evaluating/executing an
expression/statement $\mathit{stmnt}$. The remaining code to be
executed in $\mathit{cfg}'$ is $\mathit{stmnt}'$. In this way the SOS
rules define the transition relation.

To formally define configurations we need a notion of computation
state. Intuitively, each state assigns to program variables and
contract fields\footnote{\emph{State variables} in Solidity
  terminology.} their current value. It is characteristic of smart
contracts to distinguish between \emph{volatile} memory and
\emph{permanent} memory, where the former stores temporary information
produced during computation of a self-contained task (also called a
\emph{transaction}) while the latter is information that may influence
the execution of subsequent transactions and is thus stored on the
blockchain:

\begin{dfn}[Domain and State]
  The set of semantic values $\dom$ is called \emph{domain}. For each
  contract type, ADT or primitive type $\tau$ there is a domain
  $\dom^{\tau}\subseteq \dom$.  A \emph{state} is pair $\state$ of
  \emph{volatile memory} $\volatile$ and \emph{permanent} memory
  $\permanent$ where
  \begin{itemize}
  \item volatile memory is a mapping $\volatile:\Var\rightarrow\dom$
    from program variables to their domain~$\dom$;
  \item permanent memory is a mapping
    $\permanent:\domContract\rightarrow(\domField\rightarrow\dom)$,
    which assigns each contract its own persistent memory (where $\domContract:=\bigcup_{C\in\Contract}\dom^{C}$).
  \end{itemize}
\end{dfn}

We can now define a \emph{configuration}, which forms the context
wherein \smartML operates and that is modified by the execution of a
\smartML program.
    
\begin{dfn}[Configuration]
  A \emph{configuration} is a quintuple
  $$
  \config{\overbrace{\vphantom{[}c_0}^{\text{contract}}, \overbrace{\contrs}^{\text{call stack}}, \overbrace{\state}^{\text{state}}, \overbrace{\vphantom{[}\trans}^{\text{rollback permanent memory}}, \overbrace{\vphantom{[}m_0 \mapsto cnt_0}^{\text{continuation}}}
  $$
  consisting of:
  \begin{itemize}
  \item The current active contract $c_0\in\domContract$;
  \item the call stack
    $\contrs = c_1[s_1,m_1,cnt_1]\circ\cdots\circ c_n[s_n,m_n,cnt_n]$,
    where each argument triple of the current caller $c_i$ contains
    the state $s_i$ in which $c_i$ was suspended, the method $m_i$
    from where the call originated, and the remaining code $cnt_i$ to
    be executed by $c_i$;
  \item the current state $\sigma_0=(s_v^{0},s_p^{0})$;
  \item $\trans$ is a sequence of permanent memories
    $\permanent_1,\ldots,\permanent_k$; in case of a revert the system
    reverts back to the first state~$\permanent_1$ in the sequence
  \item the continuation~$m_0 \mapsto cnt_0$, i.e.~the remaining code
    $cnt_0$ to be executed next in scope of the currently active
    method~$m_0$.
  \end{itemize}
\end{dfn}

\ifthenelse{\boolean{extended}}{ Table \ref{semantics} shows selected
  SOS rules, we provide the complete semantics in
  Appendix~\ref{appendix-semantics}.}{ Table \ref{semantics} shows
  selected SOS rules, we provide the complete semantics in the
  extended version of the paper~\cite{ExtendedVersion} for the
  reviewers' convenience.  }

We start with the assignment rule~\rulename{E-Assign}. It is
applicable in a state~$(s_v^{0},s_p^{0})$ when the first statement of
a continuation is an assignment with a program variable~$x$ on the
left side and a type compatible \emph{side-effect free} expression~$e$
on the right side. The code following the assignment is matched
by~$r$. Execution of the assignment leads to an updated configuration,
whose continuation is just~$r$ with the assignment removed and its
effect reflected in the updated volatile store~$s'_v$, which is
identical to $s_v^{0}$ except for the value of program
variable~$x$. The value of~$s'_v(x)$ is equal to the value of the
assignment's right hand side~$e$ evaluated in the original
state~$(s_v^{0},s_p^{0})$.

Rule~\rulename{E-MethodCall (w/o Trans)} defines the effect of an
internal method invocation that does not open a new transaction. The
invocation~$\text{\codeinline{this}}.n(\bar{e})$ of method~$n$ on the
current contract leads to the following configuration changes:
\begin{itemize}
\item When returning without a revert/error from the call, execution
  must continue with the code after the invocation statement. Hence,
  we record the current context on the call stack. This involves
  pushing a record on the stack which is composed of
  \begin{enumerate*}[(i)]
  \item the caller~$c_0$, 
  \item the volatile memory~$s_{v}^{0}$, 
  \item the currently executed method~$m_0$,
  \item the continuation~$r$ to be executed upon return of the call
    (the program counter).
  \end{enumerate*}
  The current contract remains the active contract instance, but the
  called method~$n$ becomes now the active method, whose body is
  executed next.
\item The volatile storage~$s'_{v}$ accessible to the called
  method~$n$ consists initially only of the values passed as arguments
  ($\llbracket \bar{e}\rrbracket_{(s_v^{0},s_p^0)}$)
\item As no new transaction is opened (and the current one not
  closed), the list of transactions remains unchanged.
\end{itemize}

\begin{table}[t]
\centering
\begin{adjustbox}{width=\textwidth}
${\def\arraystretch{1.5}
    \begin{array}{c}
      \mathrule{E-Assign}{v_e = \llbracket e\rrbracket_{(s_v^0,s_p^0)}\quad e \;\text{side-effect free} \quad s'_v=s_v^0[x\leftarrow v_e] }
{(c_0,\contrs,(s_v^0,s_p^0),\trans,m_0\mapsto	x \coloneqq e; r) \leadsto (c_0,\contrs,(s'_v,s_p^0),\trans,m_0\mapsto	r)}\\
\mathrule{E-MethodCall (w/o Trans)}{s'_v=[\bar{a}\leftarrow \llbracket \bar{e}\rrbracket_{(s_v^{0},s_p^0)}]\quad \bar{e} \;\text{side-effect free} \quad n(\overline{\tau a})\{ \mathit{body}_n\}}
{\def\arraystretch{0.25}\begin{array}{l}
(c_0,\contrs,(s_v^0,s_p^0),\trans,m_0\mapsto	\mbox{\codeinline{this}}.n(\bar{e}); r) \leadsto \\
\hspace*{2.5cm} (c_0,[c_0[s_v^0,m_0,r],\contrs],(s'_v,s_p^0),\trans,n\mapsto \mathit{body}_n)
\end{array}}\\
      \mathrule{E-MethodCall (w/ Trans)}{s'_v=[\bar{a}\leftarrow \llbracket \bar{e}\rrbracket_{(s_v^0,s_p^0)}]\quad \bar{e} \;\text{side-effect free} \quad n(\overline{\tau a})\{ \mathit{body}_n\}}
{\def\arraystretch{0.25}
\begin{array}{l}
          (c_0, [\contrs],(s_v^0,s_p^0),\trans,m_0\mapsto \mbox{\codeinline{try}}\ u.n(\bar{e})\ \mbox{\codeinline{abort}}\ \{cb\}  \ \mbox{\codeinline{success}}\ \{st\}; r) \leadsto\\ 
          \hspace*{1cm}(c_u,[c_0[s_v^0,\mbox{\codeinline{try}}\ ?\ \mbox{\codeinline{abort}}\ \{cb\}  \ \mbox{\codeinline{success}}\ \{st\},r],\contrs],(s'_v,s_p^0),
        [s_p^0,\trans],n\mapsto	\mathit{body}_n)
\end{array}
}\\
\mathrule{E-MethodCall~(ReturnFromTry II)}{v=\llbracket e \rrbracket_{(s_v^0,s_p^0)}}
{\def\arraystretch{0.25}
\begin{array}{l}
\big(c_0,[c_1[s_v^1,\mbox{\codeinline{try}}\ ?\ \mbox{\codeinline{abort}}\ \{cb\}  \ \mbox{\codeinline{success}}\ \{st\},r],\contrs],
(s_v^0,s_p^0),[s_p^1,\trans],m_0\mapsto \mbox{\codeinline{throw}}\ e\big) \\
\hspace*{6.75cm}\leadsto(c_1,\contrs,(s_v^1,s_p^1),\trans,m_1\mapsto	cb(v);r)
\end{array}
}
\end{array}
}
$
\end{adjustbox}
\smallskip

\caption{Selected rules for the \smartML SOS semantics}
\label{semantics}
\end{table}

Method invocations embedded in a \codeinline{try-abort-success}
statement open a new transaction. The \codeinline{try-abort-success}
statement provides the means for appropriate error handling in case of
a failed transaction. The semantics for a method invocation that opens
a new transaction~\rulename{E-MethodCall (w/ Trans)} is similar to the
previous rule, but we have to extend the list of transactions by
recording the current permanent store $s_p^{0}$ so can revert the
state in case of an abort. The continuation put on the call stack
contains still the \codeinline{try-abort-success}, but with the actual
invocation statement replaced by an anonymous marker~$?$.

Finally, we have a look at one of the rules for returning from a
method invocation in the context of a
\codeinline{try-abort-success}~statement. We focus on the
rule~\rulename{E-MethodCall (ReturnFromTry II)} for a failed
transaction. In that case, we have to revert the permanent storage
back to the state before opening the transaction, i.e., instead of
continuing execution in~$s_p^0$, we continue with the permanent
storage~$s_p^1$. The code executed next is the body of the
\codeinline{abort}-clause, where its pattern variable~$v$ is bound to
the thrown error~$e$.

\section{Formalisation of Reentrance Safety}
\label{sec:formalisation-reentrance-safety}

We define different versions of reentrance safety for a given \smartML
contract~$C$. First we define what we mean by a reachable
configuration.

\begin{definition}[Reachable
  Configuration\label{def:reachable-definition}]
  We call a configuration~$\cfg$ \emph{reachable} (for a given smart
  contract~$C$), if it can be derived in a finite number of SOS steps
  from an initial configuration of the form
  $\config{c_0, [], (s_{v}^{0},s_p^{0}), s_p^{0}, m_0 \mapsto cnt_0}$.
\end{definition}

We say a reentrance is \emph{present} in the execution of a method~$m$
of a contract instance~$c_0$ (of type~$C$), if the invocation of~$m$
\begin{enumerate*}[(i)]
\item starts a transaction, in other words, it is not an internal
  call;
\item\label{enum:re:ext-call} contains a call to a method~$n$ of a
  different contract~$d$ as well as
\item a subsequent call to some method of $c_0$ before returning from
  the call to~$n$.
\end{enumerate*}
This is formalized as follows:

\begin{definition}[\label{def:reentrance}Rentrance]
  Given a reachable~configuration for a contract~$c_0$ of type~$C$ of
  the form $\cfg:=\defConfig$
  $\text{with\ } cs_r:=c_r[s_v^r,m_r, cnt_r]$. A \emph{reentrance is
    present} in $\cfg$ iff the formula
  \begin{align*}
    \reentrace(\cfg):=\exists i,k,j.\overbrace{\big(i\not=j \rightarrow (c_i=c_j \wedge m_i=m_j \wedge i<k<j \wedge c_k\not=c_i)\big)}^{\reentraceM(\cfg, i,k,j)}
  \end{align*}
  holds. 
\end{definition}

The following definition introduces the concept of \emph{strict
  reentrance safety} which guarantees that no reentrant calls occur
within the smart contract.

\begin{definition}[Strict Reentrance
  Safety\label{def:strict-reentrance-safety}]
  A smart contract $c_0$ of type $C$ is \emph{strictly reentrance
    safe} iff for all reachable configurations $\cfg$, the formula
  $\neg\big(\exists k,j.\reentraceM(\cfg,0,k,j)\big) \text{ holds.}$
\end{definition}

The definition above can only be satisfied by contracts that do not
invoke other contracts, thus remaining entirely
self-contained. However, if no modifications are made to fields of
reentrant contracts after a call, then reentrancy is still considered
to be \emph{safe}. Such a more liberal notion of reentrance safety can
be defined using the function $\Fields$, which returns the fields of a
contract. We call the resulting notion \textit{non-modifying
  reentrance safety}.

\begin{definition}[Non-Modifying Reentrance Safety\label{def:non-mod-reentrance-safety}]
  A smart contract~$c_0$ of type~$C$ is \emph{non-modifying reentrance
    safe} when for all reachable configurations~$\cfg$, for all~$j,k$
  such that $\reentraceM(\cfg,0,k,j)$ holds, it follows that for all
  $l>k$:
  $$c_l=c_0 \, \land \, \Fields(c_0) \cap \Fields(c_l) = \emptyset\enspace.$$
\end{definition}

Smart contracts may contain fields that are inessential for preventing
reentrancy vulnerabilities. Let us call such fields \emph{irrelevant}
and denote them as the set $\irrelevantFields$. Irrelevant fields do
not contain critical assets, such as balance and flag variables that
play a vital role in guarding against reentrancy attacks.  Unlike
balances and flags, variables in $\irrelevantFields$ are not involved in
fund management or in controlling a contract's execution flow. While
irrelevant fields may influence a contract's behavior or store data,
their modification after external calls is unlikely to introduce
reentrancy vulnerabilities. By focusing solely on relevant fields we
can establish a more liberal and practically applicable definition of
reentrancy safety.

\begin{definition}[Modifying Reentrance
  Safety\label{def:mod-irrelevant-reentrance-safety}]
  A smart contract $c_0$ of type $C$ is \emph{modifying reentrance
    safe} when for all reachable configurations $\cfg$, for all $j,k$
  such that $\reentraceM(\cfg,0,k,j)$ holds, it follows that for all
  $l>k$:
  $$c_l=c_0 \, \land \, \Fields(c_0) \cap \Fields(c_l) \subseteq \irrelevantFields\enspace.$$
\end{definition}

The set of irrelevant fields can be specified by trusted user
annotations. However, it may also be derived from specifications, for
instance, if fields are not constrained/used by invariants of a
contract or parts of the invariant are not needed for verifying a
contract's methods.



\section{Asserting Safe Reentrancy}
\label{sec:assert-safe-reentr}

To ensure safe reentrancy for \smartML contracts, we
present a type system preventing unsafe reentrancy, while
permitting provably safe reentrant calls.

\subsection{Contract Locations and Field Access}
\label{sec:typesystem2}

To implement the policy \emph{Modifying Reentrancy Safety} (see Definition~\ref{def:mod-irrelevant-reentrance-safety}), we must ensure that the contract does not access any relevant fields after an external call.
Thus, we have to collect all memory locations in the permanent memory 
(in other words, the memory locations for fields of contract instances) to 
which read and write accesses may occur within a given sequence of statements.
To represent these locations, we introduce symbolic values that refer
to contract instances. These play the role of the values assigned to
program variables or fields. Further, we need to represent memory
locations to which values can be assigned. These memory locations are
either program variables or the fields of contracts.

\begin{definition}[Contract Identities, Locations]
  The set~$\ContractIDs$ contains for each contract type
  $C\in\Contract$ infinitely many symbolic constants~$\cntId$ of type
  $C$ (disjoint from program variables) that represent a
  \emph{contract identity} (i.e., the semantic value of $\cntId$ are
  the objects in~$\dom^{C}$). We use $\ContractIDs^{\tau}$ for the set
  of all contract identities of type~$\tau$.
  We permit aliasing, i.e., two contract identity
  symbols~$\cntId_1, \cntId_2 \in \ContractIDs$ may refer to the same
  contract identity.
  A \emph{contract location} is a
  pair~$(c,f)\in \ContractIDs \times Field$.
  The set of all contract locations is called $\Location$.
  The set of all memory locations $\MemLocation$ is defined as
  $\MemLocation:=\ProgVars\cup\Location$.
\end{definition}

To determine whether a reentrancy might be problematic, tracking read
and write access to fields is mandatory. To model field accesses we
define a function
$\locs: \ContractIDs \times
(\mathsf{Statement}\cup\mathsf{Expression})\rightarrow\mathsf{2^{\Location}}$,
that collects all permanent memory locations accessed by a statement
or expression $p$ in the context of contract $c$.  Function~$\locs$ is
defined inductively on the syntactic structure of $p$:
\[
  \locs(c, p) = 
  \begin{cases}
    \{(c,f)\}\cup\locs(c,e) & \text{if } p \equiv [\this.f\;\coloneqq\;e]\\
    \{(c,f)\}& \text{if } p \equiv [\this.f]\\
    \ldots\\
    \locs(c,s_1)\cup\locs(c,s_2)&\text{if } p \equiv [s_1; s_2]\\
    \locs(c,\textbf{\textit{mbody}}(C,m))&\text{if } p \equiv [\this.m(\overline{v})],\ C \text{ type of } c\\
    \{(c,f)\mid f \in \textbf{\textit{fields}}(C)\} &\text{if } p \equiv [d.m(\overline{v})],\ d\not= \this,\ C \text{ type of } c \\
  \end{cases}
\]
\ifthenelse{\boolean{extended}}{%
  $\locs$ makes use of the lookup functions \textit{\textbf{fields}},
  \textit{\textbf{mtype}} and \textit{\textbf{mbody}}, the complete
  definitions can be found in Appendix~\ref{appendix-typesystem}.}{%
  $\locs$ makes use of the lookup functions \textit{\textbf{fields}}
  and \textit{\textbf{mbody}}, the complete definitions can be found
  in the extended version of the paper~\cite{ExtendedVersion}.  }
  
\subsection{Blocking Unsafe Reentrancy via Locks}
\label{sec:typesystem1}

The goal of the type system is to prevent reentrant calls by using a
locking mechanism. Upon invocation of a function from another
contract, or a function within the same contract that modifies its
fields, that function is considered locked, ensuring exclusive access
and preventing reentrancy vulnerabilities.

A typing judgment has the shape
$\Gamma; \Delta; \PartState; \multiset \vdashthis{this_c}{m} s
\Rrightarrow \Gamma'; \Delta';\PartState';\multiset' $ with input
context ($\Gamma; \Delta; \PartState; \multiset$), caller reference
$\this_c$, executing method $\mathtt{m}$, statement $s$ to be typed,
and output context ($\Gamma'; \Delta'; \PartState'; \multiset'$).  The
presence of a set of contract locations $\multiset$ is justified,
because an expression can change the object references that determine
reentrancy.  An empty context slot is represented by
symbol~$\nothing$.

Context slot~$\Gamma$ is a \emph{data typing environment}, mapping
program variables and fields~$x$ to their types. We write
$\Gamma, x:\tau$ for the \emph{data typing environment} $\Gamma'$ that
is equal to $\Gamma$ except that it maps~$x$ to type $\tau$.  The
possible types~$\tau$ are:
$$
  \begin{array}{rl}
    \tau & \coloneqq\  \mathtt{int} \mid \mathtt{bool} \mid \mathtt{string} \mid \mathtt{address} \mid \adt \mid \cnt \mid \stm
  \end{array}
$$
where $\cnt\in\Contract$, $\adt$ is the type name of an ADT and $\stm$
types a \textit{statement}.

The context slot~$\Delta$ contains pairs $\langle \cntId, m\rangle$,
where $\cntId$ is a contract identity and $m$ is a locked method of
the contract. Set notation is used to add and remove elements.

To improve precision, it is useful to keep track of aliasing. For this
we need bookkeeping of contract identities in memory. Partial state
functions are used to track assignments to contract-typed memory
locations. We can then use partial states $\PartState$ to compute an
over-approximation for the aliasing relation.

\begin{definition}[Partial State]
  A \emph{partial state} function
  $\PartState:\MemLocation\rightharpoonup 2^{\ContractIDs}$ maps
  memory locations to a set of contract identities. Partial states are
  undefined for memory locations that are not declared as a contract
  type.

  We write $\PartState+[\ml \mapsto K]$ for the partial state function
  that results from $\PartState$ by adding the mapping from memory
  location $\ml$ to the set of contract instances $K$.

  We define the initial partial state function
  $\PartState_{\mathsf{init}}$ that maps each memory location $\ml$ of
  contract type $\tau$ exactly to $\ContractIDs^{\tau}$.
\end{definition}

For example, $\PartState(\ml)=\{ \cntId_1, \cntId_2 \}$ means that the
value of $\ml$ is one of the contract identities $\cntId_1$ or
$\cntId_2$. Two memory locations $\ml_1, \ml_2$ are possibly aliased,
if $\PartState(\ml_1)\cap \PartState(\ml_2)\not=\emptyset$.

The context $\multiset$ is a multiset that contains all memory
locations accessed by the body of the method undergoing type
checking. We use standard multiset notation for operations on
$\multiset$. In this notation, elements in $\multiset$ take the form
$(\cntId,f)^n$, indicating that element $(\cntId,f)$ has multiplicity
$n$.


Table~\ref{type-system1} shows selected typing rules 
\ifthenelse{\boolean{extended}}{ (more are in
  Appendix~\ref{appendix-typesystem}).}{(more are in the paper's
  extended version~\cite{ExtendedVersion}).} The symbol ~$<:$ denotes the usual object
subtyping relation.

\begin{table}[t]
  $
  \begin{array}{c}
    \mathruleWW{Succ}{\begin{array}{c}\Gamma; \Delta; \PartState; \multiset\vdashthis{this_c}{m}s_1 : \stm \Rrightarrow \Gamma_1; \Delta_1; \PartState_1; \multiset_1\quad
                        \Gamma_1; \Delta_1; \PartState_1; \multiset_1 \vdashthis{this_c}{m}s_2 : \stm \Rrightarrow \Gamma_2; \Delta_2; \PartState_2;\multiset_2
                      \end{array}
    }
    {\Gamma; \Delta; \PartState; \multiset \vdashthis{this_c}{m} s_1; s_2  : \stm \Rrightarrow \Gamma_2; \Delta_2; \PartState_2; \multiset_2} \\[1ex]
    \mathruleWW{Assign}{\Gamma \vdash v :\tau \quad\tau \neq \cnt\quad \Gamma  \vdash e: \tau' \quad \tau'<: \tau}
    {\Gamma; \Delta; \PartState; \multiset \vdashthis{this_c}{m}v \coloneqq e : \stm \Rrightarrow \Gamma; \Delta; \PartState; \multiset\smallsetminus\locs(\PartState(\this_c),v \coloneqq e) }\\\\
    \mathruleWW{Assign-Cnt}{
    \begin{array}{l}
      \Gamma \vdash v : \cnt ~ \Gamma  \vdash \rhs: \cnt \quad
      (\rhs\in\MemLocation\!\Rightarrow\!\Mod=\PartState(e)) \vee (\rhs \;\text{is complex expr}\!\Rightarrow\!\Mod=\ContractIDs)
    \end{array} 
    }
    {\Gamma; \Delta; \PartState; \multiset \vdashthis{this_c}{m}v \coloneqq \rhs : \stm \Rrightarrow \Gamma; \Delta; \PartState + [v \mapsto \Mod]; \multiset\smallsetminus\locs(\PartState(\this_c),v \coloneqq \rhs)}\\\\
      \mathruleWW{If-Else}{
        \begin{array}{c}
        \Gamma\vdash e : \texttt{bool} \qquad\Gamma; \Delta; \PartState; \multiset  \vdashthis{this_c}{m}s_i : \stm \Rrightarrow \Gamma_i; \Delta_i; \PartState_i;  \multiset_i\quad\text{for } i\in\{1,2\}
        \end{array}}
        {
        \begin{array}{l}
          \Gamma; \Delta; \PartState; \multiset \vdashthis{this_c}{m}\texttt{if}\; (e)\; \{ s_1 \} \;\texttt{else}\; \{ s_2 \} : \stm 
          \Rrightarrow \\ 
          \qquad\qquad\qquad\qquad \Gamma_1\cup\Gamma_2; \Delta_1\cup\Delta_2; \PartState_1\cup\PartState_2; \multiset\smallsetminus(\multiset_1\cup\multiset_2\cup\locs(\PartState(\this_c),e))
        \end{array}
        }\\
    \mathruleWW{Call-Safe}{
    \begin{array}{c}
      \Gamma \vdash v : \cnt \quad \textit{\textbf{mtype}}(\cnt,m_v)=\overline{\tau} \longrightarrow \tau_0 \quad \Gamma \vdash \overline{u}: \overline{\tau} \\
      \Gamma, \textit{\textbf{fields}}(v); \Delta; \PartState \vdash \textbf{\textit{mbody}}(\cnt,m_v) \; \textit{ok} \quad 
      \langle\PartState(v),\textit{m}_v\rangle \cap \Delta = \nothing 
    \\  \quad \multiset \subseteq \irrelevantFields \;\lor \;(\PartState(v)=\PartState(\this_c)\;\land\;\locs(\PartState(v),\textbf{\textit{mbody}}(\cnt,m_v))\subseteq \irrelevantFields)
    \end{array}
    }{\Gamma; \Delta; \PartState; \multiset \vdashthis{this_c}{m}\textit{v.m}_v(\overline{u}) : \stm \Rrightarrow \Gamma; \Delta; \PartState; \multiset\smallsetminus\locs(\PartState(\this_c),v.m_v(\overline{u}))}\\
    \mathruleWW{Call}{
    \begin{array}{c}
      \Gamma \vdash v : \cnt \quad  \textit{\textbf{mtype}}(\cnt,m_v)=\overline{\tau} \longrightarrow \tau_0 \quad \Gamma \vdash \overline{u}: \overline{\tau}\\
      \Gamma, \textit{\textbf{fields}}(v); \Delta\cup\{\langle\PartState(\this_c),\textit{m}\rangle\}; \PartState \vdash \textbf{\textit{mbody}}(\cnt,m_v) \; \textit{ok}  \quad\langle\PartState(v),\textit{m}_v\rangle \cap \Delta = \nothing
    \end{array}
    }
    {\Gamma; \Delta; \PartState; \multiset\vdashthis{this_c}{m}\textit{v.m}_v(\overline{u}) : \stm \Rrightarrow \Gamma; \Delta\cup\{\langle\PartState(v),m_v\rangle\}; \PartState; \multiset\smallsetminus\locs(\PartState(\this_c),v.m_v(\overline{u}))}\\\\
    \mathruleWW{Mth-Ok}{
    \begin{array}{c}
      c = \mathtt{contract}~C~\mathtt{extends}~D\; \{\ldots\}\;\quad
      \textit{\textbf{mtype}}(D,m)=\overline{\tau} \longrightarrow \tau_0 \quad
      \Gamma \vdash \overline{v} : \overline{\tau} \quad
      \multiset = \locs(\PartState(\this_c),s)\\
      \Gamma, \overline{v} : \overline{\tau}; \Delta; \PartState + [\overline{v}\mapsto\overline{\ContractIDs}]; \multiset \vdashthis{this_c}{m} s : \stm \Rrightarrow \Gamma'; \Delta'; \PartState'; \multiset'\\
    \end{array}
    }{\Gamma; \Delta; \PartState \vdash m(\overline{v})\{s\} \; ok }
    \\
    \sosRuleType{Cnt-Ok}{%
    \begin{array}{c} \textit{\textbf{fields}}(D) = \overline{g} : \overline{\tau}_g \qquad
      \mathtt{constructor}(\overline{g} : \overline{\tau}_g, \overline{f} : \overline{\tau}_f)\{\mathtt{super}(\overline{g});\; \this.\overline{f} = \overline{f} \}\\
      \mathit{ctx} \vdash m_1(\overline{v}_1)\{s_1\}  \; \textit{ok};
      \quad\cdots\quad
      \mathit{ctx} \vdash m_n(\overline{v}_n)\{s_n\} \; \textit{ok}\\
      \cntId\; \mathit{fresh} \text{ contract identity and } \mathit{ctx}:=\overline{g} : \overline{\tau}_g,\overline{f} : \overline{\tau}_f; \nothing; \PartState_{\mathsf{init}} + [\this_c \mapsto \cntId]
    \end{array}
    }
    {
    \vdash \mathtt{contract}~C~\mathtt{extends}~D\; \{\; \overline{f} : \overline{T}_f;\; \mathtt{constructor}(\overline{f},\overline{g});\; m_1;\ldots; m_n \}\; \textit{ok}
    }
  \end{array}
  $\\[1.5ex]
  \caption{Selected typing rules for \smartML}
  \label{type-system1}
\end{table}    

To verify that a smart contract~$C$ has safe reentrancy, we begin with
rule \rulename{Cnt-Ok}. This rule creates a new contract
identity~$\cntId$ for $C$, and initiates type checking for each of its
methods.
Next, rule \rulename{Mth-Ok} validates the well-typedness of each of
$C$'s methods.  Here, the multiset $\multiset$ is initialized for the
remaining type-checking process to the result of applying $\locs$ to
the method's body.

Before explaining the statement-level rules, we highlight that for
each rule, the output context for~$\multiset$ is determined by
\emph{excluding} the memory locations accessed by the involved
statement, which are calculated by function~$\locs$.

We split the rule for assignment statements into two cases depending
on whether the assigned variable is of contract type, because
we need to track the locations of contract references. Hence, in rule \rulename{Assign-Cnt}, the output context
$\PartState$ is updated depending on whether~$e$ is a memory location or
a complex expression. For a memory location, we update~$v$ to 
$\PartState(e)$, otherwise, we safely approximate the range of values by 
the set of all contract identities of corresponding type.  

Rule \rulename{Succ} is straightforward. The outputs
of~\rulename{If-Else} are the union of the outputs of each
branch. This prevents reentrancy, but we possibly over-approximate the
contracts' current locations when contract assignments are involved.

Rules \rulename{Call-Safe} and \rulename{Call} prevent reentrancy.
Rule~\rulename{Call-Safe} checks the \emph{safe} method calls. There
are two scenarios when a call is considered safe:
\begin{enumerate*}[label=(\Roman*)]
\item A call to a method within the same contract that leaves the
  contract's relevant field variables unaltered and thus satisfies
  $\locs(\PartState(v),\textbf{\textit{mbody}}(\cnt,m_v)) \subseteq
  \irrelevantFields$. This check is crucial for preventing
  cross-function reentrancy.
\item One can ensure that all checks or updates on the contract's
  relevant fields were completed before initiating the call itself.
  This is achieved by checking that the multiset $\multiset$ contains
  only irrelevant fields. In this way, we are sure that no pending
  access to the contract's relevant fields occur after the call.
\end{enumerate*}
Both call rules ensure that the targeted method is currently unlocked
by examining whether~$\Delta$ contains a possible alias of the caller
$v$ by checking condition
$\langle \PartState(v),m_v\rangle$.\footnote{\label{fn:partState}Due
  to the possibility of a contract being associated with multiple
  locations, by a slight abuse of notation we identify
  $\langle \PartState(v),m_v\rangle$ with
  ${\langle \cntId_1,m_v\rangle,\ldots,\langle \cntId_n,m_v\rangle}$,
  where $\PartState(v) = \{\cntId_1,\ldots,\cntId_n$\}.}  The
\rulename{Call} rule adds the known aliases of $v$ to the encountered
memory locations, whereas rule \rulename{Call-Safe} does not, because
it proved accesses to be safe.

\setcounter{lemma}{0}
We show adequacy of the type system through two fundamental
properties: type preservation, which ensures that the types are
maintained throughout evaluation, and progress, which guarantees that
well-typed programs do not get stuck during
execution. \ifthenelse{\boolean{extended}}{}{The proofs can be found
  in~\cite{ExtendedVersion}.}

Recall from Section~\ref{sec:semantics} that
$\cfg[\mathit{s}]\leadsto\cfg'[\mathit{s}']$ means a statement $s$ in
an SOS configuration $\cfg$ is reduced to a statement $s'$ in
configuration $\cfg'$.

\begin{thm}[Type Preservation]
  Let $c = \mathtt{contract}~C~\mathtt{extends}~D\; \{\ldots\}$ be a
  contract with $\vdash c\ \sf{ok}$, $s$ a statement of $c$. If
  $\Gamma; \Delta; \PartState; \multiset \vdash s : \stm \Rrightarrow
  \Gamma'; \Delta';\PartState';\multiset' $ and
  $\mathit{cfg}[\mathit{s}] \leadsto \mathit{cfg}'[\mathit{s}']$, then
  $\exists \, \Gamma_1, \Delta_1$ such that
  $\Gamma \subseteq \Gamma_1$, $\Delta \subseteq \Delta_1$ and
  $\Gamma_1; \Delta_1; \PartState; \multiset \vdash s' : \stm
  \Rrightarrow \Gamma'; \Delta';\PartState';\multiset'.$
\end{thm}
\ifthenelse{\boolean{extended}}{
\begin{proof}
  Given our semantics, we have that \( s \) and \( s' \) are, respectively, \( s \equiv \{s_1; r\} \) and \( s' \equiv \{s_1'; r\} \). Therefore, according to rule \rulename{Succ}, to prove the theorem, we need to show that \( \Gamma_1; \Delta_1; \PartState_1; \multiset_1 \vdash s_1' : \stm \Rrightarrow \Gamma_1^{\ast}; \Delta_1^{\ast}; \PartState_1^{\ast}; \multiset_1^{\ast} \), given that \( \Gamma; \Delta; \PartState; \multiset \vdash s_1 : \stm \Rrightarrow \Gamma^{\ast}; \Delta^{\ast}; \PartState^{\ast}; \multiset^{\ast} \).
  Moreover, since the statements \( s \) type well and \( \Gamma \subseteq \Gamma_1 \), we already know that \( \Gamma^{\ast}; \Delta^{\ast}; \PartState^{\ast}; \multiset^{\ast} \vdash r : \stm \Rrightarrow \Gamma_1'; \Delta_1'; \PartState_1'; \multiset_1' \).
  
  The proof proceeds by induction on the application of the transition rules.

\begin{description}
  \item [Case \rulename{E-Assign}:] By assumption, $\conf{x\coloneqq e; r} \leadsto \confp{r}$ holds.
  Since from the hypothesis $\Gamma; \Delta; \PartState; \multiset  \vdashthis{this_c}{m} \{x\coloneqq e; r \}: \stm \Rrightarrow \Gamma'; \Delta'; \PartState';  \multiset'$, we can apply the \rulename{Succ} rule and from the premises we derive that $\Gamma_1, \Delta_1; \PartState_1; \multiset_1  \vdashthis{this_c}{m} r : \stm \Rrightarrow \Gamma'; \Delta'; \PartState';  \multiset'$, with $\Gamma_1 = \Gamma$.

  \item [Case \rulename{E-If-Then-True}:] By assumption, we know that $\conf{\texttt{if}\, (e)\, \{ s_1 \} \, \texttt{else}\, \{ s_2 \} ; r}\leadsto \confp{s_1;r}$ and $\Gamma; \Delta; \PartState; \multiset \vdashthis{this_c}{m}\texttt{if}\, (e)\, \{ s_1 \} \,\texttt{else}\, \{ s_2 \} : \stm 
  \Rrightarrow \Gamma'; \Delta'; \PartState'; \multiset'$ hold. From the premises of rule \rulename{If-Else}, it follows that $\exists \Gamma_1 = \Gamma$ such that $\Gamma, \Delta; \PartState; \multiset  \vdashthis{this_c}{m} s_1 : \stm \Rrightarrow \Gamma'; \Delta'; \PartState';  \multiset'$.
  \item [Case \rulename{E-If-Then-False}:] Same as \rulename{If-Then-False}.
  \item [Case \rulename{E-WhileLoopCnt}:] By assumption $\conf{\texttt{while} (e) \{ s \} ; r}\leadsto \confp{s; r}$ and, from the premises of rule \rulename{While}, $\mathsf{fixpoint}(\Gamma_i; \Delta_i; \PartState_i; \multiset_i \vdashthis{this_c}{m}s : \stm \Rrightarrow \Gamma_{i+1}; \Delta_{i+1}; \PartState_{i+1}; \multiset_{i+1})$ hold. It follows that for each iteration $\Gamma_1 = \Gamma_i$ and $\Gamma_1; \Delta_1; \PartState_1; \multiset_1 \vdashthis{this_c}{m} s : \stm \Rrightarrow \Gamma'_{1}; \Delta'_{1}; \PartState'_{1}; \multiset'_{1}$.
  \item [Case \rulename{E-WhileLoopExit}:] Follows from the \rulename{WhileLoopCnt} case and the \rulename{Succ} rule.
  \item [Case \rulename{E-Let}:] By assumption $\conf{\texttt{let}\, x\coloneqq \rhs\, \texttt{in}\, s; r}$ transits to $\confp{s; r}$ and $\Gamma; \Delta; \PartState; \multiset \vdashthis{this_c}{m} \texttt{let}\, x \coloneqq \rhs \,\texttt{in}\, s : \stm \Rrightarrow \Gamma'; \Delta'; \PartState';\multiset'$ holds.
 Thanks to the premise of the rule \rulename{Let}, it follows that $\Gamma_1; \Delta_1; \PartState_1; \multiset_1 \vdashthis{this_c}{m} \texttt{let } x \coloneqq \rhs \texttt{ in } s : \stm \Rrightarrow \Gamma'_1; \Delta'_1; \PartState'_1;\multiset'_1$ with $\Gamma_1 = \Gamma, x : \tau$. 
  \item [Case \rulename{E-MethodCall (w/ Trans)}:] By assumption $\conf{u.n(\bar{e}); r} \leadsto \confp{\mathit{body}_n}$ holds. In this case, we do not need to distinguish between a call and a safe call, because both rules check the body of the called method. In particular, from the premises of \rulename{Call} and \rulename{Call-Safe}, $\Gamma \cup \textit{\textbf{fields}}(u); \Delta_1; \PartState \vdash \textbf{\textit{mbody}}(u,n) \; \sf{ok}$ holds. Moreover, the rule \rulename{Mth-Ok} checks the statements of the body of the method, since $\textit{body}_n \equiv \{s_1; \ldots; s_n\}$ ($\Gamma\cup\textit{\textbf{fields}}(u), \overline{e} : \overline{\tau}; \Delta_1; \PartState_1; \multiset_1 \vdashthis{this_c}{m} s : \stm \Rrightarrow \Gamma_1'; \Delta_1'; \PartState_1'; \multiset_1'$). Thus, with $\Gamma_1 = \Gamma \cup \textit{\textbf{fields}}(u) \cup [\overline{e} : \overline{\tau}]$, we have $\Gamma_1; \Delta_1; \PartState_1; \multiset_1 \vdashthis{this_c}{m} \textit{body}_n : \stm \Rrightarrow \Gamma'_1; \Delta'_1; \PartState'_1;\multiset'_1$.
  \item [Case \rulename{E-MethodCall (w/o Trans)}:] Same as \rulename{MethodCall (w/ Trans)}.
  \item [Case \rulename{E-MethodCall (ReturnFromTry I-II-III)}:] Follow from rule \rulename{Return} and case \rulename{MethodCall (w Trans)}. 
  \item [Case \rulename{E-Try-Catch}:] Same as \rulename{MethodCall (ReturnFromTry II)}.

\end{description}
\end{proof}}{}
\begin{thm}[Progress]
  If
  $\,\Gamma; \Delta; \PartState; \multiset \vdash s : \stm
  \Rrightarrow \Gamma'; \Delta';\PartState';\multiset' $ for
  statement~$s$, then for any configuration $\mathit{cfg}$ there
  exists $s'$ such that
  $\mathit{cfg}[\mathit{s}] \leadsto \mathit{cfg}'[\mathit{s}']$.
  \end{thm}
\ifthenelse{\boolean{extended}}{
\begin{proof}
The proof can be constructed using induction based on the application of the type system rules.
\end{proof}
}{}
The next theorem ensures reentrancy prevention, and thus that \smartML contracts are \emph{modifying reentrant safe}.
\begin{thm}[Reentrancy Security]
  Any smart contract that can be typed as
  $\vdash\mathtt{contract}~C~\mathtt{extends}~D\; \{\ldots\}\;
  \sf{ok}$ is modifying reentrance safe.
\end{thm}
\ifthenelse{\boolean{extended}}{
\begin{proof}
The proof is by contradiction. 
Assume there exists a smart contract $c_0 : C$ such that $\vdash c_0 \; \sf{ok}$, but we assue that $c_0$ is \textbf{not} non-modifying reentrance safe.
    By Definition \ref{def:reentrance}, a reentrance is present in $\mathsf{cfg}$ if there exist integers $i$, $k$, and $j$ such that $i \neq j$, $c_i = c_j$, $m_i = m_j$, $i < k < j$, and $c_k \neq c_i$.
    Since $c_0$ is not non-modifying reentrance safe, there exists an $l > k$ where $c_l = c_0$ and $\Fields(c_0) \cap \Fields(cnt_l) \nsubseteq \irrelevantFields$, contrary to the condition stated in Definition \ref{def:non-mod-reentrance-safety}.
    Furthermore, by hypothesis, we know that $\vdash C \; \sf{ok}$, meaning that the contract types well as well as its methods and their bodies. 
    For this reason, to check if it is possible that the condition $\Fields(c_0) \cap \Fields(cnt_l) \nsubseteq \irrelevantFields$ is satisfied, meaning that the fields of the contract $c_0$ are modified in the continuation $cnt_l$, it is essential to examine the rules \rulename{Call-Safe} and \rulename{Call}. 
\begin{itemize}
\item    The rule \rulename{Call-Safe} allows a call either if the relevant fields of the contract are not modified after it ($\multiset \subseteq \irrelevantFields$) or it is an internall call that leaves the
    contract's field variables unaltered. That means that the relevant fields of the callee contract must remain unaltered after the called method. Thus, it follows that  $ \nexists\, l>k$ such that $\Fields(c_0) \cap \Fields(cnt_l) \nsubseteq \irrelevantFields$. 

\item   The \rulename{Call} rule permits calls to external methods whose function bodies are not specified, allowing modifications to fields within the method. However, if a function satisfies this condition, the callee method is included in the set \(\Delta\).
This check ensures that the callee method cannot be invoked again during its execution. Consequently, it prevents the contract \(c_l\) (where \(c_l = c_0\) by hypothesis) from being called after invoking \(c_k\), as this would be prohibited by the condition \(\langle\PartState(c_l), \textit{m}_l\rangle \cap \Delta = \emptyset\) of the \rulename{Call} rule.
\end{itemize}
    Therefore, we can conclude that
    $$ \nexists\, l>k \text{ such that } \Fields(c_0) \cap \Fields(cnt_l) \nsubseteq \irrelevantFields$$
    Then, our initial assumption that there exists a smart contract $c_0 : C$ such that $\vdash C \; \sf{ok}$, but $c_0$ is not non-modifying reentrance safe, must be false.
    Hence, we have proven that every smart contract $c_0 : C$ satisfying $\vdash C \; \sf{ok}$ is non-modifying reentrance safe.

\end{proof}}{}


\section{Reentrancy Mitigation}
\label{evaluation}

\subsection{Case Study}
\label{sec:case-study}

This section illustrates the power of \smartML's type system for
preventing reentrancy attacks. We analyze two \smartML contracts to
show how the type system enforces secure execution flow, effectively
eliminating the possibility of
reentrancy. Listing~\ref{lst:store-contract-code} presents the
\codeinline{Store} contract, while Listing~\ref{lst:attacker-code} presents
\ifthenelse{\boolean{extended}}{}{\begin{wrapfigure}[20]{r}{.38\textwidth} \vspace{-2.9em}}
\begin{lstlisting}[style=sml-color,columns=fullflexible, xleftmargin=0.75em, numberstyle=\scriptsize, numbersep=3pt,  tabsize=1,basicstyle=\sffamily\footnotesize,label={lst:adt-in-smartml},caption={ListInt ADT},numbers=left,captionpos=b]
datatype ListInt { @\label{adt:init}@
 constructor { 
   nil | cons(int v, ListInt tail)
 }
 int indexOf(ListInt l, int n) {
  switch (l) {
   case nil: return -1;
   default: 
    if (l.v == n) { return 0; } 
    else {
     int idx = indexOf(l.tail,n);
     switch (idx) {
      case -1: return -1;
      default: return idx + 1;
     }
   } } }
  ListInt add(ListInt l, int e) { 
    return cons(e, l); 
  }
 } @\label{adt:end}@
\end{lstlisting}	
\ifthenelse{\boolean{extended}}{}{\end{wrapfigure}} the code of the
\codeinline{Attacker}. The code features a cross-function reentrancy
attack, where the attacker attempts to withdraw more funds than
permitted by invoking the \codeinline{transfer} function within its
\codeinline{receive} function.  We describe the code of the
listings. Listing~\ref{lst:adt-in-smartml} defines a list of integers
as an ADT (\codeinline{ListAddress} is analogous), excluding standard
setter and getter methods.

The \codeinline{Store} contract is in
Listing~\ref{lst:store-contract-code}, except the constructor which
follows the usual pattern.  The \codeinline{withdraw} function checks
the balance of the caller and performs the internal call
\codeinline{this.transfer(bal,index)}. An example of an external call
with resource consumption is in line~\ref{try:init} of the function
\codeinline{transfer}.  Lines~\ref{try:init}--\ref{try:end} show
transaction handling using a \codeinline{try-abort-success} statement,
allowing developers detailed control over nested transactions in
\smartML.  For lack of space, we do not show the deposit function,
which adds the resource specified by the caller to the correct
address.  In Listing~\ref{lst:attacker-code}, the \codeinline{attack}
function deposits and then tries to withdraw funds from the
store. When funds are received, the \codeinline{receive} function
increases the balance and recursively triggers another
withdrawal. This process aims to drain the \codeinline{Store}'s funds
by repeatedly invoking \codeinline{transfer}.
\begin{figure*}[h!]
\hfil\begin{minipage}[h]{.51\textwidth}
\begin{lstlisting}[style=sml-color,columns=fullflexible, numberstyle=\scriptsize, tabsize=2,basicstyle=\sffamily\footnotesize,label={lst:store-contract-code},caption={Store contract},numbers=left,captionpos=b,numbersep=4pt]
contract Store { @\label{cnt:init}@
 ListAddress addr; ListInt balances;
 function withdraw() {
  int bal = 0;
  int idx = addr.indexOf(addr,sender);
  if (idx != -1) { bal = balances.get(idx); }
  assert(bal > 0);  
  this.transfer(bal,idx);      
 }
 bool function transfer(int amount,int idx) {
  try @\color{violet}\bfseries sender\color{black}@$bal.receive();@\label{try:init}@
  abort { return false; }
  success {
   balances.set(idx,0);
   return true;
  }@\label{try:end}@
 } } @\label{cnt:end}@
\end{lstlisting}
\end{minipage}\hfil
\begin{minipage}[h]{.375\textwidth}
\begin{lstlisting}[style=sml-color,escapechar=\#,columns=fullflexible, numberstyle=\scriptsize, tabsize=2,basicstyle=\sffamily\footnotesize,label={lst:attacker-code},caption={Attacker},numbers=left,captionpos=b,numbersep=4pt]
contract Attacker {
 int balance; 
 Store s;
 constructor() {
  this.balance = 1;
  this.s = new Store();
 }
 
 function attack() {
  s$balance.deposit();
  s$0.withdraw();
 }
 
 function receive() {
  balance=balance+#\(\langle\)#amount#\(\rangle\)#;
  s$0.transfer();
 } }
\end{lstlisting}
\end{minipage}
\end{figure*}

Our type system effectively blocks the repetitive call by leveraging
the
set~$\Delta$. Specifically, when the type system evaluates whether the
\codeinline{transfer} function of the \codeinline{Store} contract can
be invoked within the \codeinline{receive} function of the Attacker,
the derivation process encounters a failure. This occurs because
\codeinline{this}$_{\codeinlineSub{Store}}$ and \codeinline{store} are
aliased, meaning they share the same partial state. Consequently, the
type system detects a conflict as it attempts to verify that the set
$\Delta$ does not include
$\langle\PartState(\codeinline{store}),\codeinline{transfer}\rangle$
(recalling the notational convention in
footnote~\ref{fn:partState}). This conflict indicates a potential
reentrancy vulnerability, leading the type system to block the
operation and thereby ensuring the security of the contract execution
(see Figure~\ref{proof-type-system}).

\begin{figure}[h]
\newcommand{\thisStore}{\text{\codeinline{this}$_{\codeinlineSub{Store}}$}}
\newcommand{\thisAtt}{\text{\codeinline{this}$_{\codeinlineSub{Att}}$}}
  \centering
  \begin{adjustbox}{width=\textwidth}
\begin{prooftree}[rule margin=1ex]
      \hypo{\xmark}
      \infer1[]{ \{\langle \PartState(\thisAtt),\codeinline{attack}\rangle,\langle \PartState(\thisStore),\codeinline{withdraw}\rangle,\langle \PartState(\thisStore),\codeinline{transfer}\rangle\}\cap\langle\PartState(\codeinline{store}),\codeinline{transfer}\rangle = \nothing}
      \infer1[]{ \{\langle \PartState(\thisAtt),\codeinline{attack}\rangle,\langle \PartState(\thisStore),\codeinline{withdraw}\rangle,\langle \PartState(\thisStore),\codeinline{transfer}\rangle\}\vdashthis{this_{Att}}{receive} \codeinline{store}.\codeinline{transfer()} } 
      \infer1[]{\ldots} 
 
      \infer1[]{\begin{array}{l}\{\langle \PartState(\thisAtt),\codeinline{attack}\rangle,\langle \PartState(\thisStore),\codeinline{withdraw}\rangle\}\vdashthis{this_{Store}}{transfer} \codeinline{att.receive()} \Rrightarrow\\[1ex]
   \hspace*{15em} \{\langle \PartState(\thisAtt),\codeinline{attack}\rangle,\langle \PartState(\thisStore),\codeinline{transfer}\rangle,\langle \PartState(\thisStore),\codeinline{transfer}\rangle\}\end{array}}    
      \infer1[]{\ldots} 

      \infer1[]{ \{\langle \PartState(\thisAtt),\codeinline{attack}\rangle\}\vdashthis{this_{Store}}{withdraw} \thisStore\codeinline{.transfer(amount,index)} \Rrightarrow \{\langle \PartState(\thisAtt),\codeinline{attack}\rangle,\langle \PartState(\thisStore),\codeinline{withdraw}\rangle\}}    
      \infer1[]{\ldots} 
      \infer1[]{ \nothing \vdashthis{this_{Att}}{\codeinline{attack}} \codeinline{store.withdraw()} \Rrightarrow \{\langle \PartState(\thisAtt),\codeinline{attack}\rangle\}}  
      \infer1[]{\ldots}         
    \end{prooftree}
   \end{adjustbox}

  \caption{Type derivation for the example (relevant checks and changes to $\Delta$ only)}
  \label{proof-type-system}
\end{figure}

\subsection{Safe Reentrancy}

While a complete ban on reentrancy seems like a simple solution, it is
overly restrictive and reduces the functionality and
interoperability of smart contracts.
For this reason, our type system has been carefully designed as a
safeguard against unsafe reentrant calls while permitting those
considered to be secure. A key feature is the ability to assess
whether a call to an external contract occurs after all necessary
checks and updates to the fields have been executed. When such a call
satisfies the non-interference condition, it is considered safe. Such
calls are not added to the set~$\Delta$ of locked method calls. Thus,
this approach to designing our type system serves a dual purpose: it
effectively prevents reentrancy attacks while enabling the execution
of safe calls, finding a middle ground that avoids unnecessary
restrictions.

\section{Related Work}
\label{related-works}

Various smart contract languages address reentrancy vulnerabilities
using different methods. Scilla~\cite{scilla,SergeyNJ0TH19} is an
intermediate language for verified smart contracts, relying on
communicating automata and Coq for proving contract properties.
Scilla avoids reentrancy by removing the call-and-return paradigm in
contract interactions. However, their approach is not compositional in
the sense that it fails to block cross-contract reentrancy.
Obsidian~\cite{obsidian} and Flint~\cite{flint} are two smart contract
languages that enhance contract behavior comprehensibility with
typestate integration. Obsidian includes a dynamic check for
object-level reentrancy, whereas Flint lacks a reentrancy check. Like
Scilla, both languages lack compositionality, i.e.~fail to block
cross-contract reentrancy, even though both incorporate a linear asset
concept to prevent certain attacks.

SeRIF~\cite{serif} detects reentrancy based on a trusted-untrusted
computation model using a type system with trust labels for secure
information flow. It spots cross-contract reentrancy without blocking
every reentrant call. However, we avoid a control flow type system,
maintaining flexibility and expressiveness without imposing
constraints on program structure.
Nomos~\cite{nomos} adopts a security enforcement strategy grounded in
session types. The linearity of session types does not fully address
reentrancy, hence, the paper employs the resources monitored by these
session types as a safeguard. This approach ensures that attackers
cannot gain authorization to invoke a contract that is currently in
use, eliminating all forms of reentrancy, even safe reentrancy. In
contrast, our approach permits safe tail reentrancy calls.
SolType~\cite{soltype} is a refinement type system for Solidity that
prevents over- and under-flows in smart contracts. While the type
system is very powerful concerning arithmetic bugs in smart contracts,
it does not provide a safety guarantee against reentrancy.

There are several static analysis tools for smart contracts:
Oyente~\cite{oyente} is a bug finding tool with no soundness
guarantees, based on symbolic execution.
While symbolic execution is a powerful generic technique for
discovering bugs, it does not guarantee to explore all program paths
(resulting in false negatives).  \textsc{Securify}~\cite{securify} is
a tool for analyzing Ethereum smart contracts and its analysis
consists in two steps. First, it symbolically analyzes the contract's
dependency graph to extract precise semantic information from the
code. Then, it checks compliance and violation
patterns.
Both tools, focus on one or two contracts, and thus, sequences and
interleavings of function calls from multiple contracts are often
ignored. In contrast, our approach guarantees security against
cross-contract reentrancy attacks.

Several tools employ formal verification to analyze contracts, like
VERISOL~\cite{verisol} which is a highly automated formal verifier for
Solidity. It not only generates proofs, but also identifies
counterexamples, ensuring smart contracts align with a state machine
model including of access control policies.
Solythesis~\cite{solythesis} is a source-to-source Solidity compiler
that takes a smart contract and a user-specified invariant as its
input and produces an instrumented contract that rejects all
transactions that violate the invariant.
These tools focus on single contract safety, so they lack the ability
of compositional verification.

\begin{figure}[t]
  \begin{center}
    \includegraphics[width=.8\textwidth]{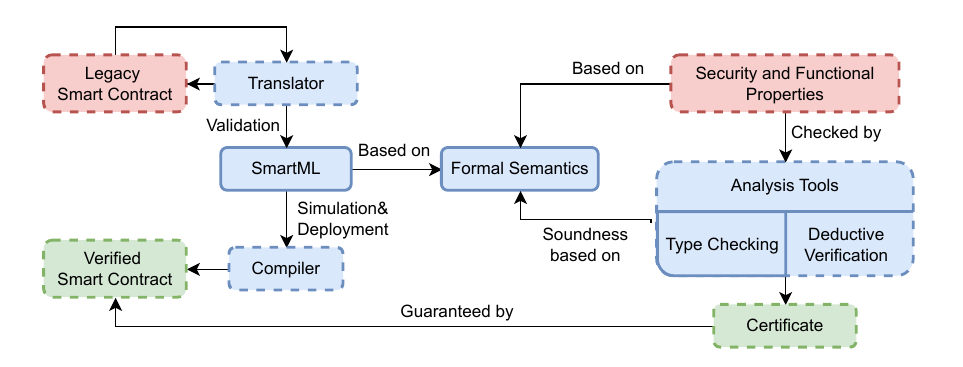}
  \end{center}
  \vspace*{-1ex}
  \caption{\label{fig:smartML}Overview of the \smartML framework}
\end{figure}

\section{Conclusion and Future Work}
\label{conclusions}

We presented a language-independent modeling framework for smart
contracts, see Figure~\ref{fig:smartML} (elements with a solid border
are discussed in this paper).  {\smartML} offers a comprehensive
approach to formally specifying and verifying smart contracts,
mitigating the inherent complexities and vulnerabilities that can
expose security-critical assets to unforeseen attacks.  A
platform-independent modeling language complements the state-of-art by
providing a structured and abstract representation of contracts. This
facilitates understanding and analysis.  To be fully
platform-independent, we are currently developing a translator from
existing smart contract languages to {\smartML} (and back).

A formal operational semantics and a type system for safe reentrancy
checks further establishes a robust foundation for expressing and
verifying functional correctness as well as security properties of
smart contracts.  A deductive verification and a static analysis tool,
with the aim of proving the absence of relevant classes of security
vulnerabilities and functional correctness of the smart contracts, are
ongoing work.
This paper's contributions pave the way for future advancements in
blockchain research, emphasizing the importance of addressing security
concerns to fully unlock the potential of distributed ledger
technology.  

\bibliographystyle{plainurl}
\bibliography{references}

\newpage
\ifthenelse{\boolean{extended}}{
\appendix
\section{Full {\smartML} Semantics}
\label{appendix-semantics}

The full operational semantics for \smartML is given in Table
\ref{semantics-complete1} and Table \ref{semantics-complete2}.
\begin{table}[h]
  \centering
  \begin{adjustbox}{width=\textwidth}
  ${\def\arraystretch{1.5}
      \begin{array}{c}
    \mathrule{E-Assign}{v_e = \llbracket e\rrbracket_{(s_v^0,s_p^0)}\quad e \;\text{side-effect free} \quad s'_v=s_v^0[x\leftarrow v_e] }
    {(c_0,\contrs,(s_v^0,s_p^0),\trans,m_0\mapsto	x \coloneqq e; r) \leadsto (c_0,\contrs,(s'_v,s_p^0),\trans,m_0\mapsto	r)}\\

\mathrule{E-MethodCall (w/o Trans)}{s'_v=[\bar{a}\leftarrow \llbracket \bar{e}\rrbracket_{(s_v^{0},s_p^0)}]\quad \bar{e} \;\text{side-effect free} \quad n(\overline{\tau a})\{ \mathit{body}_n\}}
{\def\arraystretch{0.25}\begin{array}{l}
(c_0,\contrs,(s_v^0,s_p^0),\trans,m_0\mapsto	\mbox{\codeinline{this}}.n(\bar{e}); r) \leadsto \\
\hspace*{2.5cm} (c_0,[c_0[s_v^0,m_0,r],\contrs],(s'_v,s_p^0),\trans,n\mapsto \mathit{body}_n)
\end{array}}\\
      \mathrule{E-MethodCall (w/ Trans)}{s'_v=[\bar{a}\leftarrow \llbracket \bar{e}\rrbracket_{(s_v^0,s_p^0)}]\quad \bar{e} \;\text{side-effect free} \quad n(\overline{\tau a})\{ \mathit{body}_n\}}
{\def\arraystretch{0.25}
\begin{array}{l}
          (c_0, [\contrs],(s_v^0,s_p^0),\trans,m_0\mapsto \mbox{\codeinline{try}}\ u.n(\bar{e})\ \mbox{\codeinline{abort}}\ \{cb\}  \ \mbox{\codeinline{success}}\ \{st\}; r) \leadsto\\ 
          \hspace*{1cm}(c_u,[c_0[s_v^0,\mbox{\codeinline{try}}\ ?\ \mbox{\codeinline{abort}}\ \{cb\}  \ \mbox{\codeinline{success}}\ \{st\},r],\contrs],(s'_v,s_p^0),
        [s_p^0,\trans],n\mapsto	\mathit{body}_n)
\end{array}
}\\
\mathrule{E-MethodCallAssign (w/ Trans)}{s'_v=[\bar{a}\leftarrow \llbracket \bar{e}\rrbracket_{(s_v^0,s_p^0)}]\quad \bar{e} \;\text{side-effect free} \quad n(\overline{\tau a})\{ \mathit{body}_n\}}
{\def\arraystretch{0.25}
\begin{array}{l}
    (c_0,\contrs,(s_v^0,s_p^0),\trans,m_0\mapsto \mbox{\codeinline{try}}\ x\coloneqq u.n(\bar{e})\ \mbox{\codeinline{abort}}\ \{cb\}  \ \mbox{\codeinline{success}}\ \{st\}; r) \leadsto\\ 
    \hspace*{1cm} (c_u,[c_0[s_v^0,\mbox{\codeinline{try}}\ x\coloneqq ?\ \mbox{\codeinline{abort}}\ \{cb\}  \ \mbox{\codeinline{success}}\ \{st\},r],\contrs],(s'_v,s_p^0),[s_p^0,\trans],n\mapsto	\mathit{body}_n)
  \end{array}
}\\
\mathrule{E-MethodCall~(ReturnFromTry I)}{\llbracket e \rrbracket_{(s_v^0,s_p^0)}}
    {\def\arraystretch{0.25}
    \begin{array}{l}
    (c_0,[c_1[s_v^1,\mbox{\codeinline{try}}\ ?\ \mbox{\codeinline{abort}}\ \{cb\}  \ \mbox{\codeinline{success}}\ \{st\},r],\contrs],(s_v^0,s_p^0),[s_p^1,\trans],m_0\mapsto	 \mbox{\codeinline{return}})\leadsto\\
    \hspace*{8.75cm} (c_1,\contrs,(s_v^1,s_p^0),\trans,m_1\mapsto	st; r)
    \end{array}
    }\\
    \mathrule{E-MethodCall~(ReturnFromTry II)}{v=\llbracket e \rrbracket_{(s_v^0,s_p^0)}}
{\def\arraystretch{0.25}
\begin{array}{l}
\big(c_0,[c_1[s_v^1,\mbox{\codeinline{try}}\ ?\ \mbox{\codeinline{abort}}\ \{cb\}  \ \mbox{\codeinline{success}}\ \{st\},r],\contrs],
(s_v^0,s_p^0),[s_p^1,\trans],m_0\mapsto \mbox{\codeinline{throw}}\ e\big)\leadsto \\
\hspace*{8.75cm}(c_1,\contrs,(s_v^1,s_p^1),\trans,m_1\mapsto	cb(v);r)
\end{array}
}\\
    \mathrule{E-MethodCall~(ReturnFromTry III)}{v=\llbracket e \rrbracket_{(s_v^0,s_p^0)}}
    {\def\arraystretch{0.25}
    \begin{array}{l}
    (c_0,[c_1[s_v^1, \mbox{\codeinline{try}}\ x\coloneqq~?\ \mbox{\codeinline{abort}}\ \{cb\}  \ \mbox{\codeinline{success}}\ \{st\},r],\contrs],(s_v^0,s_p^0),[s_p^1,\trans],m_0\mapsto	\mbox{\codeinline{return}}\ e)\leadsto \\
    \hspace*{8.75cm}(c_1,\contrs,(s_v^1,s_p^1),\trans,m_1\mapsto	x\coloneqq v; st ;r)
    \end{array}
    }\\
    \mathrule{E-If-Then-True}{\llbracket cond \rrbracket_{(s_v^0,s_p^0)}=\text{true}}
    {(c_0,\contrs,(s_v^0,s_p^0),\trans,m_0\mapsto	 \mbox{\codeinline{if}}\ (cond)  \{s\} \ \mbox{\codeinline{else}}\ \{s'\}; r)\leadsto (c_0,\contrs,(s_v^0,s_p^0),\trans,m_0\mapsto	s;r)
    }\\
    \mathrule{E-If-Then-False}{\llbracket cond \rrbracket_{(s_v^0,s_p^0)}=\text{false}}
    {(c_0,\contrs,(s_v^0,s_p^0),\trans,m_0\mapsto	 \mbox{\codeinline{if}}\ (cond) \{s\} \ \mbox{\codeinline{else}}\ \{s'\}; r)\leadsto (c_0,\contrs,(s_v^0,s_p^0),\trans,m_0\mapsto	s';r)
    }\\
    
  \end{array}
  }
  $
  \end{adjustbox}
  \vspace{0.5cm}
  \caption{The \smartML semantics}
  \label{semantics-complete1}
  \end{table}
\clearpage

  \begin{table}[t]
    \centering
    \begin{adjustbox}{width=\textwidth}
    ${\def\arraystretch{1.5}
        \begin{array}{c}
  \mathrule{E-WhileLoopCnt}{\llbracket cond \rrbracket_{(s_v^0,s_p^0)}=\text{true}}
    {(c_0,\contrs,(s_v^0,s_p^0),\trans,m_0\mapsto	 \mbox{\codeinline{while}}\ (cond) \{ s \} ; r)\leadsto (c_0,\contrs,(s_v^0,s_p^0),\trans,m_0\mapsto	s; \mbox{\codeinline{while}}\ (cond) \{ s \};r)
    }\\
    \mathrule{E-WhileLoopExit}{\llbracket cond \rrbracket_{(s_v^0,s_p^0)}=\text{false}}
    {(c_0,\contrs,(s_v^0,s_p^0),\trans,m_0\mapsto	 \mbox{\codeinline{while}}\ (cond) \{ s \}; r)\leadsto (c_0,\contrs,(s_v^0,s_p^0),\trans,m_0 \mapsto r)
    }
    \\
    \mathrule{E-Let}{v_e = \llbracket e\rrbracket_{(s_v^0,s_p^0)}\quad e \;\text{side-effect free} \quad s_{stat}=s_v^0[x\leftarrow v_e] }
    {(c_0,\contrs,(s_v^0,s_p^0),\trans,m_0\mapsto	 \mbox{\codeinline{let}}\ x \coloneqq e \ \mbox{\codeinline{in}}\ stat; r)\leadsto (c_0,\contrs,(s_{stat},s_p^0),\trans,m_0\mapsto stat; r)
    }\\
  \end{array}
  }
  $
  \end{adjustbox}
  \vspace{0.5cm}
  \caption{The \smartML semantics (cont.)}
  \label{semantics-complete2}
  \end{table}

    \clearpage
\section{Full \smartML Type System}
\label{appendix-typesystem}

The full \smartML type system rules are presented in Table \ref{type-system-appendix3-cnt} and in Table \ref{type-system-appendix3}.
\begin{table}[h!]
  \centering
\begin{tabular}{c} 
    \multicolumn{1}{l}{\textbf{Lookup Functions}} \\\\
    \sosRuleTypeBis{
          \begin{array}{c}
              \texttt{contract } C \texttt{ extends } D\; \{\; \overline{f} : \overline{T}_f;\; \texttt{constructor}(\overline{f},\overline{g});\; \overline{M} \}\\
              \textit{\textbf{fields}}(D) = \overline{g} : \overline{\tau}_g 
          \end{array}}{\textit{\textbf{fields}}(C) = \overline{f} : \overline{\tau}_f; \overline{g} : \overline{\tau}_g}
       \\\\
      \sosRuleTypeBis{
      \begin{array}{c}
          \texttt{contract } C \texttt{ extends } D\; \{\; \overline{f} : \overline{T}_f;\; \texttt{constructor}(\overline{f},\overline{g});\; \overline{M} \}\\
          \tau_0 m(\overline{\tau}\;\overline{x}) \in \overline{M}
      \end{array}
       }{\textit{\textbf{mtype}}(C,m) = \overline{\tau} \longrightarrow \tau_0}
      \\\\
                
  \sosRuleTypeBis{
      \begin{array}{c}
          \texttt{contract } C \texttt{ extends } D\; \{\; \overline{f} : \overline{T}_f;\; \texttt{constructor}(\overline{f},\overline{g});\; \overline{M} \}\\
          \tau_0 m(\overline{\tau}\;\overline{x})\{\overline{s}\} \in \overline{M}
      \end{array}
       }{\textit{\textbf{mbody}}(C,m) = \overline{s}}
      \\\\\hline\\
    \end{tabular}

    \begin{tabular}{c} 
    \multicolumn{1}{l}{\textbf{Value Typing}} \\\\
    $\begin{array}{c}
      \sosRuleTypeNew{Var}{\Gamma(x)=\tau}{\Gamma \vdash x:\tau}
\qquad
\sosRuleTypeNew{Int}{}{\vdash n:\texttt{int}}\qquad
\sosRuleTypeNew{True}{}{\vdash \texttt{true}:\texttt{bool}}
\qquad
\sosRuleTypeNew{False}{}{\vdash \texttt{false}:\texttt{bool}}\\\\
\sosRuleTypeNew{Op}{\Gamma \vdash x_1 : \tau \quad\Gamma \vdash x_2 : \tau}{\Gamma \vdash x_1 \;\texttt{op}\; x_2 : \tau}\quad
\sosRuleTypeNew{BOp}{\Gamma \vdash x_1 : \tau \quad\Gamma \vdash x_2 : \tau}{\Gamma \vdash x_1 \;\texttt{bop}\; x_2 : \texttt{bool}}
\end{array}$
\\\\\hline\\
\end{tabular}
  \begin{tabular}{c} 
    \multicolumn{1}{l}{\textbf{Contract Typing}} \\\\         
$\sosRuleType{Mth-Ok}{
      \begin{array}{c}
        c = \mathtt{contract}~C~\mathtt{extends}~D\; \{\;\ldots\}\;\\
        \textit{\textbf{mtype}}(D,m)=\overline{\tau} \longrightarrow \tau_0\\
        \Gamma \vdash \overline{v} : \overline{\tau}\\
        \multiset = \locs(\PartState(\this_c),s)\\
        \Gamma, \overline{v} : \overline{\tau}; \Delta; \PartState + [\overline{v}\mapsto\overline{\ContractIDs}]; \multiset \vdashthis{this_c}{m} s : \stm \Rrightarrow \Gamma'; \Delta'; \PartState'; \multiset'\\
      \end{array}
      }{\Gamma; \Delta; \PartState \vdash m(\overline{v})\{s\} \; \sf{ok} }
$
    \\\\

      $  \begin{array}{c}
        \sosRuleType{Cnt-Ok}{\begin{array}{c} \textit{\textbf{fields}}(D) = \overline{g} : \overline{\tau}_g \\ 
          \mathtt{constructor}(\overline{g} : \overline{\tau}_g, \overline{f} : \overline{\tau}_f)\{\mathtt{super}(\overline{g});\; \this.\overline{f} = \overline{f} \}\\
          \cntId\; \mathit{fresh}, \text{ contract identity} \\
          \overline{g} : \overline{\tau}_g,\overline{f} : \overline{\tau}_f; \nothing; \PartState_{\mathsf{init}}+[\this_c \mapsto \cntId] \vdash m_1(\overline{v}_1)\{s_1\}  \; \sf{ok};\\
          \ldots\\
          \overline{g} : \overline{\tau}_g,\overline{f} : \overline{\tau}_f; \nothing; \PartState_{\mathsf{init}} + [\this_c \mapsto \cntId] \vdash m_n(\overline{v}_n)\{s_n\} \; \sf{ok}\\
        \end{array}
}
{
\vdash \mathtt{contract}~C~\mathtt{ext.}~D\; \{\; \overline{f} : \overline{T}_f;\; \mathtt{const.}(\overline{f},\overline{g});\; m_1;\ldots; m_n \}\; \sf{ok}
}
  
\end{array}$
\\\\\hline\\

\end{tabular}
\caption{Rules for the type system $\vdashthis{this_c}{m}$ }
\label{type-system-appendix3-cnt}
\end{table}

\begin{table}[h]
  \begin{adjustbox}{width=\textwidth}
\begin{tabular}{c} 
    \multicolumn{1}{l}{\textbf{Core Expressions Typing}} \\\\
$ \begin{array}{c}
  \sosRuleType{Succ}{\begin{array}{c}\Gamma; \Delta; \PartState; \multiset\vdashthis{this_c}{m}s_1 : \stm \Rrightarrow \Gamma_1; \Delta_1; \PartState_1; \multiset_1\qquad
    \Gamma_1; \Delta_1; \PartState_1; \multiset_1 \vdashthis{this_c}{m}s_2 : \stm \Rrightarrow \Gamma_2; \Delta_2; \PartState_2;\multiset_2
  \end{array}
}
{\Gamma; \Delta; \PartState; \multiset \vdashthis{this_c}{m}\{ s_1; s_2 \} : \stm \Rrightarrow \Gamma_2; \Delta_2; \PartState_2; \multiset_2} \\\\

  \sosRuleType{Let}{\Gamma \vdash e : \tau \quad \Gamma, x : \tau ; \Delta; \PartState; \multiset \vdashthis{this_c}{m}s : \stm \Rrightarrow \Gamma_1; \Delta_1; \PartState_1;\multiset_1}{\Gamma; \Delta; \PartState; \multiset \vdashthis{this_c}{m}\texttt{let }x\coloneqq e \texttt{ in } s : \stm \Rrightarrow \Gamma_1; \Delta_1; \PartState_1; \multiset_1 \smallsetminus \locs(\PartState(\this_c,e))}\\\\

\sosRuleType{Assign}{\Gamma \vdash v :\tau \qquad\tau \neq \cnt\qquad \Gamma  \vdash e: \tau' \qquad \tau'<: \tau}
{\Gamma; \Delta; \PartState; \multiset \vdashthis{this_c}{m}v \coloneqq e : \stm \Rrightarrow \Gamma; \Delta; \PartState; \multiset\smallsetminus\locs(\PartState(\this_c),v \coloneqq e) }\\\\

\sosRuleType{Assign-Cnt}{\Gamma \vdash v : \cnt \qquad \Gamma  \vdash e: \cnt \qquad (e\in\MemLocation \Rightarrow \Mod=\PartState(e)) \vee (e \;\text{is complex expression} \;\Rightarrow \Mod=\ContractIDs)}
{\Gamma; \Delta; \PartState; \multiset \vdashthis{this_c}{m}v \coloneqq e : \stm \Rrightarrow \Gamma; \Delta; \PartState + [v \mapsto \Mod]; \multiset\smallsetminus\locs(\PartState(\this_c),v \coloneqq e)}\\\\ 
  \sosRuleType{While}{
  \begin{array}{c} 
    \Gamma_0 \vdash e : \texttt{bool} \quad \mathsf{fixpoint}(\Gamma_i; \Delta_i; \PartState_i; \multiset_i \vdashthis{this_c}{m}s : \stm \Rrightarrow \Gamma_{i+1}; \Delta_{i+1}; \PartState_{i+1}; \multiset_{i+1})\\
  \Gamma^{\ast}; \Delta^{\ast}; \PartState^{\ast}; \multiset^{\ast} \text{ are the output contexts of the fixpoint}
  \end{array}
  }
  {\Gamma_0; \Delta_0; \PartState_0; \multiset_0\vdashthis{this_c}{m} \texttt{while } e \;\{\; s \;\} : \stm \Rrightarrow \Gamma^{\ast}; \Delta^{\ast}; \PartState^{\ast}; \multiset^{\ast}\smallsetminus\locs(\PartState(\this_c),e)}\\\\
  \sosRuleType{If-Else}{
      \begin{array}{c}
      \Gamma\vdash e : \texttt{bool} \quad\Gamma; \Delta; \PartState; \multiset  \vdashthis{this_c}{m}s_1 : \stm \Rrightarrow \Gamma_1; \Delta_1; \PartState_1;  \multiset_1\qquad
      \Gamma; \Delta; \PartState; \multiset\vdashthis{this_c}{m}s_2: \stm \Rrightarrow \Gamma_2; \Delta_2; \PartState_2; \multiset_2
      \end{array}}
      {
      \Gamma; \Delta; \PartState; \multiset \vdashthis{this_c}{m}\texttt{if}\; (e)\; s_1 \;\texttt{else}\; s_2 : \stm 
      \Rrightarrow \Gamma_1\cup\Gamma_2; \Delta_1\cup\Delta_2; \PartState_1\cup\PartState_2; \multiset\smallsetminus(\multiset_1\cup\multiset_2\cup\locs(\PartState(\this_c),e))
      }\\\\
      \sosRuleType{Try-Abort}{
    \begin{array}{c}
        \Gamma; \Delta; \PartState; \multiset \vdashthis{this_c}{m}s_0 : \stm \Rrightarrow \Gamma_0; \Delta_0; \PartState_0;  \multiset_0
    \qquad 
        \Gamma_0; \Delta_0; \PartState_0; \multiset_0  \vdashthis{this_c}{m}s_1 : \stm \Rrightarrow \Gamma_1; \Delta_1; \PartState_1; \multiset_1
    \\
    \Gamma_0; \Delta_0; \PartState_0; \multiset_0\vdashthis{this_c}{m}s_2: \stm \Rrightarrow \Gamma_2; \Delta_2; \PartState_2;   \multiset_2
    \end{array}}
    {
    \Gamma; \Delta; \PartState; \multiset \vdashthis{this_c}{m}\texttt{try}\; s_0\;\texttt{abort}\; s_1 \;\texttt{success}\; s_2 : \stm 
    \Rrightarrow \Gamma_0\cup\Gamma_1\cup\Gamma_2; \Delta_0\cup\Delta_1\cup\Delta_2; \PartState_0\cup\PartState_1\cup\PartState_2; \multiset\smallsetminus(\multiset_0\cup\multiset_1\cup\multiset_2)
    }\\\\
           \sosRuleType{Return}
    {\Gamma \vdash e : \tau \quad \tau <: \textbf{\textit{mtype}}(C,m)} 
    {\Gamma; \Delta; \PartState; \multiset \vdashthis{this_c}{m}\texttt{return } e : \stm \Rrightarrow \Gamma; \Delta; \PartState; \multiset\smallsetminus\locs(\PartState(\this_c),e)}\\\\
    \sosRuleType{Assert}{\Gamma\vdash e : \texttt{bool}}
    {\Gamma; \Delta; \PartState;\multiset \vdashthis{this_c}{m}\texttt{assert} (e): \stm \Rrightarrow \Gamma; \Delta; \PartState; \multiset\smallsetminus\locs(\PartState(\this_c),e)}\\\\
    \mathruleWW{Call-Safe}{
      \begin{array}{c}
        \Gamma \vdash v : \cnt \quad \textit{\textbf{mtype}}(\cnt,m_v)=\overline{\tau} \longrightarrow \tau_0 \quad \Gamma \vdash \overline{u}: \overline{\tau} \\
        \Gamma, \textit{\textbf{fields}}(v); \Delta; \PartState \vdash \textbf{\textit{mbody}}(\cnt,m_v) \; \textit{ok} \qquad
        \langle\PartState(v),\textit{m}_v\rangle \cap \Delta = \nothing 
        \\  \quad \multiset \subseteq \irrelevantFields \;\lor \;(\PartState(v)=\PartState(\this_c)\;\land\;\locs(\PartState(v),\textbf{\textit{mbody}}(\cnt,m_v))\subseteq \irrelevantFields)
      \end{array}
      }{\Gamma; \Delta; \PartState; \multiset \vdashthis{this_c}{m}\textit{v.m}_v(\overline{u}) : \stm \Rrightarrow \Gamma; \Delta; \PartState; \multiset\smallsetminus\locs(\PartState(\this_c),v.m_v(\overline{u}))}\\
      \mathruleWW{Call}{
      \begin{array}{c}
        \Gamma \vdash v : \cnt \quad  \textit{\textbf{mtype}}(\cnt,m_v)=\overline{\tau} \longrightarrow \tau_0 \quad \Gamma \vdash \overline{u}: \overline{\tau} 
        \\
       \Gamma, \textit{\textbf{fields}}(v); \Delta\cup\{\langle\PartState(\this_c),\textit{m}\rangle\}; \PartState \vdash \textbf{\textit{mbody}}(\cnt,m_v) \; \textit{ok}  \quad\langle\PartState(v),\textit{m}_v\rangle \cap \Delta = \nothing
      \end{array}
      }
      {\Gamma; \Delta; \PartState; \multiset\vdashthis{this_c}{m}\textit{v.m}_v(\overline{u}) : \stm \Rrightarrow \Gamma; \Delta\cup\{\langle\PartState(v),m_v\rangle\}; \PartState; \multiset\smallsetminus\locs(\PartState(\this_c),v.m_v(\overline{u}))}\\\\
 \\
        \end{array}
      $
        \end{tabular}
      \end{adjustbox} 
\caption{Rules for the type system $\vdashthis{this_c}{m}$ (continuation from previous page)}
    \label{type-system-appendix3}        
\end{table}}{}
\end{document}